\documentclass[aps,pre,showpacs,amsfonts,floatfix, twocolumn]{revtex4}

\usepackage{graphics}
\usepackage{placeins}
\usepackage{graphicx}
\usepackage{amssymb}
\usepackage{amsmath}
\usepackage{hyperref}
\usepackage{color}

\date{\today}

\newcommand{\eq}[1]{(\ref{eq:#1})}
\newcommand{\fig}[1]{FIG.~\ref{fig:#1}}

\begin{document}

\title{Stochastic slowdown in evolutionary processes}
\author{Philipp M.~Altrock}
\email{altrock@evolbio.mpg.de}
\author{Chaitanya S.~Gokhale}
\author{Arne Traulsen}
\affiliation{Emmy--Noether Group for Evolutionary Dynamics, Department of Evolutionary Ecology,
Max-Planck-Institute for Evolutionary Biology, August-Thienemann-Str.\ 2, D--24306 Pl\"on, Germany\\
}
\date{\today}


\begin{abstract}
We examine birth--death processes with state dependent transition probabilities and at least one absorbing boundary. 
In evolution, this describes selection acting on two different types in a finite population where reproductive events occur successively. 
If the two types have equal fitness the system performs a random walk.  
If one type has a fitness advantage it is favored by selection, which introduces a bias (asymmetry) in the transition probabilities. 
How long does it take until advantageous mutants have invaded and taken over? 
Surprisingly, we find that the average time of such a process can increase, even if the mutant type always has a fitness advantage.  
We discuss this finding for the Moran process and develop a simplified model which allows a more intuitive understanding.  
We show that this effect can occur for weak but non--vanishing bias (selection) in the state dependent transition rates and infer the scaling with system size. 
We also address the Wright--Fisher model commonly used in population genetics, which shows that this stochastic slowdown is not restricted to birth--death processes. 
\end{abstract}

\maketitle

\section{Introducion}\label{sec:intro}

Birth--death processes belong to the most simple stochastic models and are applied in a variety of fields 
\cite{kampen:1997xg,Gardiner04,feller:1968bv,karlin:1975xg,moran:1962ef,goel:1974aa}. 
In physics these processes are connected e.g.~to the study of one--dimensional classical diffusion in disordered media,  
anomalous transport, and molecular motors \cite{Landauer1987prb,Fisher1988aa,hongler:2008pl,reimann:2002pr}. 
In evolutionary biology, birth--death processes are commonly applied to model the evolution of traits with different reproductive fitness that are under natural selection 
\cite{moran:1962ef,ewens:2004qe}. 
In the context of evolutionary game theory, this particular class of Markov chains has been used to model the spreading of successful strategies in a population of small size 
\cite{szabo:2002te,antal:2006aa,nowak:2004pw,helbing:1993aa,Schuster:2003mi,szabo:2007aa,claussen:2008aa,Berr:2009prl,roca:2009aa}. 
Naturally, the limit of weak selection is considered to be important in biology. 
It describes situations in which the effects of payoff differences are small,
such that the evolutionary dynamics are mainly driven by random fluctuations.
While this approach has a long standing history in population genetics \cite{kimura:1968aa,ohta:2002aa},
in the context of evolutionary game dynamics it has  been introduced only recently \cite{nowak:2004pw}. 
Often, from the discrete stochastic process a continuous limit or diffusion approximation is motivated, 
where typically the impact of the relevant parameters and time scales can be studied more easily \cite{ewens:2004qe,kaniovski:2000ee,traulsen:2005hp,doering:2005mm}. 
Here, we consider the Moran process from theoretical population genetics and related processes. 
We address the speed of evolution when a resident population is taken over by mutants that are more fit. 
Under the low mutation rates that typically occur in biology, a mutant either goes extinct or takes over the population 
before another mutation arises. 
Thus, for many purposes it is sufficient to address the evolution of two types in a one-dimensional system.  

In the following, we first recall general properties of birth--death processes (Sec.~\ref{sec:general}) and then address asymmetry in the transition probabilities (Sec.~\ref{sec:bias}).
In Sec.~\ref{sec:WF}, we then consider a more general Markov process to highlight that our main finding is not a special property of birth--death processes.

\section{State dependent birth--death process}\label{sec:general}

A one--dimensional birth--death process in position $i$ can move to $i-1$ or $i+1$ with probabilities $T_i^-$ and $T_i^+$. 
With probability $1-T_i^--T_i^+$, the process stays in state $i$. 
We assume $T_0^{\pm}=T_N^{\pm}=0$, such that $i=0$ and $i=N$ are absorbing states. 
In discrete time, the probability to reach boundary $N$ in $t$ steps, starting from any $i$, obeys the master equation \cite{goel:1974aa}
\begin{align}\label{eq:Master01}
\begin{split}
	P_i^N(t)\,=\,&\left(1-T_i^+-T_i^-\right)P_i^N(t-1)\\
			&+T_i^-P_{i-1}^N(t-1)+T_i^+P_{i+1}^N(t-1).
\end{split}
\end{align}
The stationary conditional $n^{\text{th}}$ moment of $P_i^N(t)$ is given by 
\begin{align}\label{eq:nMoment}
	\left( \phi_i^N  \right)^{-1}\sum_{t=0}^{\infty}t^n\,P_i^N(t).
\end{align}
The normalization constant, $\phi_i^N=\sum_{t=0}^{\infty}P_i^N(t)$, is the probability that the process gets absorbed at boundary $N$, 
called fixation probability in population genetics. 
For $\phi_i^N$ a recursion is obtained from Eq.~\eq{Master01}, 
$\phi^N_{i}=(1-T_i^+-T_i^-)\phi^N_{i}+T_i^-\phi^N_{i-1}+T_i^+\phi^N_{i+1}$.
With the boundary conditions $\phi_0^N=0$ and $\phi_N^N=1$, the solution reads \cite{karlin:1975xg}
\begin{align}\label{eq:FixProb01}
\begin{split}
		\phi_i^N=\frac
		{
		1+\sum_{k=1}^{i-1}\prod_{m=1}^{k}\frac{T_m^-}{T_m^+}
		}
		{
		1+\sum_{k=1}^{N-1}\prod_{m=1}^{k}\frac{T_m^-}{T_m^+}
		}.
\end{split}
\end{align} 
A measure for the duration of the process is the conditional mean time to absorption (average fixation time) $\tau_i^{N}$, i.e.~the first moment of $P_i^{N}(t)$. 
This gives the average number of time steps until one of the two absorbing states is reached, starting from any $i$ \cite{Landauer1987prb,antal:2006aa}. 
A recursion for $\tau_i^N$ is obtained by multiplying each side of Eq.~\eq{Master01} with $t$ and summing over all $t$ \cite{goel:1974aa}, 
which yields $\phi_i^N\,\tau_i^N=\,\left(1-T_i^+-T_i^-\right)\phi_i^N\,\tau_i^N+T_i^-\phi_{i-1}^N(\tau_{i-1}^N+1)+T_i^+\phi_{i+1}^N(\tau_{i+1}^N+1)$. 
A similar recursion can be found for the conditional mean exit time $\tau_i^0$, 
such that the mean life time of the process amounts to $\tau_i^0+\tau_i^N$.  
Solving recursively with the boundary conditions $\tau_0^N=0 $ and $\tau_N^N=0$, leads to the conditional mean time to reach state $N$, starting from $i=1$,  
\begin{align}
\label{eq:CtimeOne}
	\tau_1^N = \sum\limits_{k=1}^{N-1}\,\sum\limits_{l=1}^{k}\,\frac{\phi_l^N}{T_l^+}\prod_{m=l+1}^{k}\frac{T_m^-}{T_m^+}.
\end{align}
One common example for a birth-death process with absorbing states $0$ and $N$ is the homogenous random walk, $T_i^\pm=c \leq 1/2$ for $0 <i <N$ and $T_0^\pm = T_N^\pm =0$ \cite{redner:2001bo}. 
This leads to $\phi_i^N=i/N$ and $\tau_1^N=(N^2-1)/(6c)$. 
The reference case of population genetics is neutral evolution, where the symmetric transition probabilities are state dependent, $T_i^\pm=i(N-i)/N^2$. 
This results in $\phi_i^N=i/N$ and $\tau_1^N=N(N-1)$ \cite{moran:1962ef,ewens:2004qe}.

\section{Biased transition probabilities}\label{sec:bias}

In this section, we examine how the state dependent transition probabilities influence the conditional mean exit time. 
We consider processes in which a parameter $\beta$ continuously introduces a bias towards moving into one direction: 
For $\beta =0$ the transition probabilities are symmetric, $T_i^+ = T_i^-$, but for $\beta>0$, an asymmetry arises, $T_i^+ \geq T_i^-$. 
In evolutionary dynamics, $\beta$ is usually referred to as the intensity of selection. 
It governs the selective advantage (or disadvantage) of mutants in a wild--type population of finite size.  
Intuitively, it is clear that the time $\tau_1^N$ does not depend trivially on $\beta$, cf. Eq.\ \eq{CtimeOne}.
With increasing $\beta$, the probability $\phi_i^N$ increases, but both $1/T_i^+ $ and  ${T_i^-}/{T_i^+}$ decrease in our setup. 
Thus, the average time $\tau_1^N$ can increase or decrease with $\beta$. 
In other words, despite increasing the tendency to move in the direction of a given boundary in each state, the conditional average time until this boundary is reached can still increase. 

In the Moran process, an individual selected for reproduction proportional to fitness produces identical offspring that replaces a randomly selected individual from the population.  
We consider the evolution of two types $A$ and $B$ in a finite population of size $N$. 
Type $A$ (with fitness $f_A$) is usually referred to as the mutant type, $B$ (with fitness $f_B$) is called the wild--type. 
Let $i$ be the number of individuals of type $A$, such that $N-i$ is the number of $B$ individuals. 
In general, the transition probabilities are
\begin{align}
\begin{split}
	&T_i^+=  \frac{i\,f_A }{i f_A + (N-i)f_B}   \frac{N-i}{N} \\
	&T_i^-=  \frac{(N-i)\,f_B  }{i f_A + (N-i)f_B}   \frac{i}{N} 
\end{split}
\end{align}
In the following, we discuss different choices of $f_A$ and $f_B$, as well as closely related, but simplified asymmetric transition rates.

\subsection{Constant fitness}

In the simplest case, the fitness of mutants is constant and does not depend on their abundance \cite{ewens:2004qe}. 
In our model, this can be parametrized as $f_A=1+\beta$ and $f_B=1-\beta$.  
In this case, the fixation probability of a single mutant is \cite{ewens:2004qe}
\begin{equation}
\phi_1^{N}=(1-\gamma)/(1-\gamma^{N}), 
\label{stdfix}
\end{equation}
where $\gamma=(1-\beta)/(1+\beta)$. 
Up to linear order in $\beta$ we have 
$\phi_1^N\approx N^{-1} + \beta (N-1)N^{-1}$. 
The larger the fitness advantage, the more likely the evolutionary takeover. 
For stronger selection ($\beta>0$) an advantageous mutant is expected to fixate faster compared to neutral ($\beta=0$).

\subsection{Linear density dependence}

\begin{figure}[t]
\includegraphics[angle=0,width=0.475\textwidth]{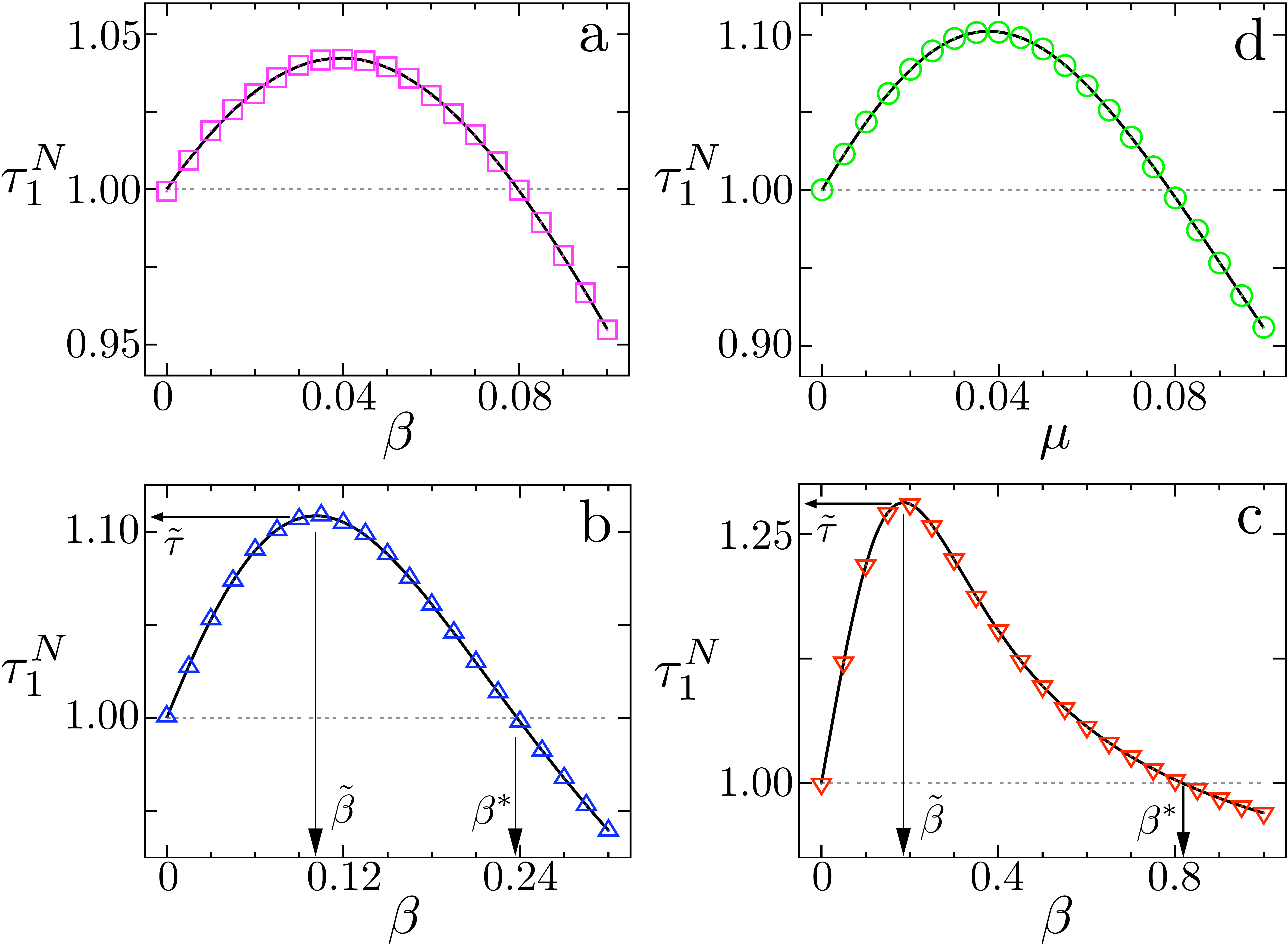}
\caption{
(Color online) 
The conditional mean exit time $\tau_1^N/\tau_1^N(0)$ (normalized) as a function of the bias (selection intensity) $\beta$, or the mutation rate $\mu$, 
for the four different models discussed in the main text. 
Symbols are simulations, lines show Eq.~\eq{CtimeOne}. 
({\bf a}) Moran process with $a=-0.1$ and $b=2$, see Eq.~\eq{Moran01}.  
({\bf b}) Parabolic--step process with $i^{\ast}=11$, Eq.~\eq{parabolstep}. 
({\bf c}) Constant--step process with $i^{\ast}=9$ and $c=0.5$, Eq.~\eq{TranP03}. 
({\bf d}) Birth--death process with directed mutations, Eqs.~\eq{Tmutplus} and \eq{Tmutminus}.
The quantities $\tilde\tau$, $\tilde\beta$, and $\beta^{\ast}$ 
indicate the maximal realtive increase of $\tau_1^N$, the according bias parameter, 
and the non--trivial value of $\beta$ where $\tau_1^N=\tau_1^N(0)$, respectively (also compare \fig{scaling}). 
The system size is $N=20$ in all panels, averages taken over $10^7$ realizations. 
}
\label{fig:hump}
\end{figure}

In general, the fitness of the two types will depend on their abundance. 
For example, the fitness $f$ of each type can change linearly with $i$, $ f_{A} = 1+ \beta\,(a\,i+b)$ and $ f_{B} = 1- \beta\,(a\,i+b)$. 
The bias $\beta$ is bound such that fitness never becomes negative. 
Then, the transition probabilities are
\begin{align}\label{eq:Moran01}
	T_i^\pm=  \frac{1\pm\beta\,(a\,i+b) }{N - \beta (a \, i + b) (N-2i)}   \frac{i \left(N-i\right)}{N}.
\end{align}
We have $T_0^\pm=T_N^\pm=0$, such that both boundaries are absorbing \cite{nowak:2004pw,taylor:2004wv}. 
For $a<0$ and $a N +b>0$, type $A$ is always fitter than type $B$, $f_{A} > f_{B}$, 
but the conditional mean exit time $\tau_1^N$ is larger than neutral in a certain parameter range, compare \fig{hump} (a). 
In this case, a mutant that is fitter than the rest of the population needs more time to take over the population than a less fit mutant 
-- intuitively, this should not be the case. 
The linear approximation of $\tau_1^N$ for $\beta \ll N^{-1}$ (weak selection) reads  
\begin{align}\label{eq:LinearCFT}
\begin{split}
	\tau_1^N \approx N(N-1) -a\,\frac{N^2(N^2-3N+2)}{18}\,\beta,
\end{split}
\end{align}
see \cite{taylor:2006jt,altrock:2009nj}. 
Note that the linear approximation of the conditional mean exit time depends only on the parameter $a$, but not on $b$, which holds for any system size. 
Hence, for small bias $\beta$ and $a<0$, the conditional average time grows with increasing $\beta$. 
This is an effect from state dependent fitness in finite populations, as it cannot occur for $a=0$. 

The ratio $T_i^-/T_i^+$ is a measure of the stochastic flow. 
Stochastic slowdown can occur if this ratio changes with the position (abundance of $A$) $i$, leading to an asymmetry. 
When $\beta$ becomes larger, $\tau_1^N$ decreases again with $\beta$, which is the strong selection behavior one would expect, compare \fig{hump} (a).

\subsection{Step--like asymmetry}\label{ssec:simple}

\begin{figure}[t]
\includegraphics[angle=0,width=0.485\textwidth]{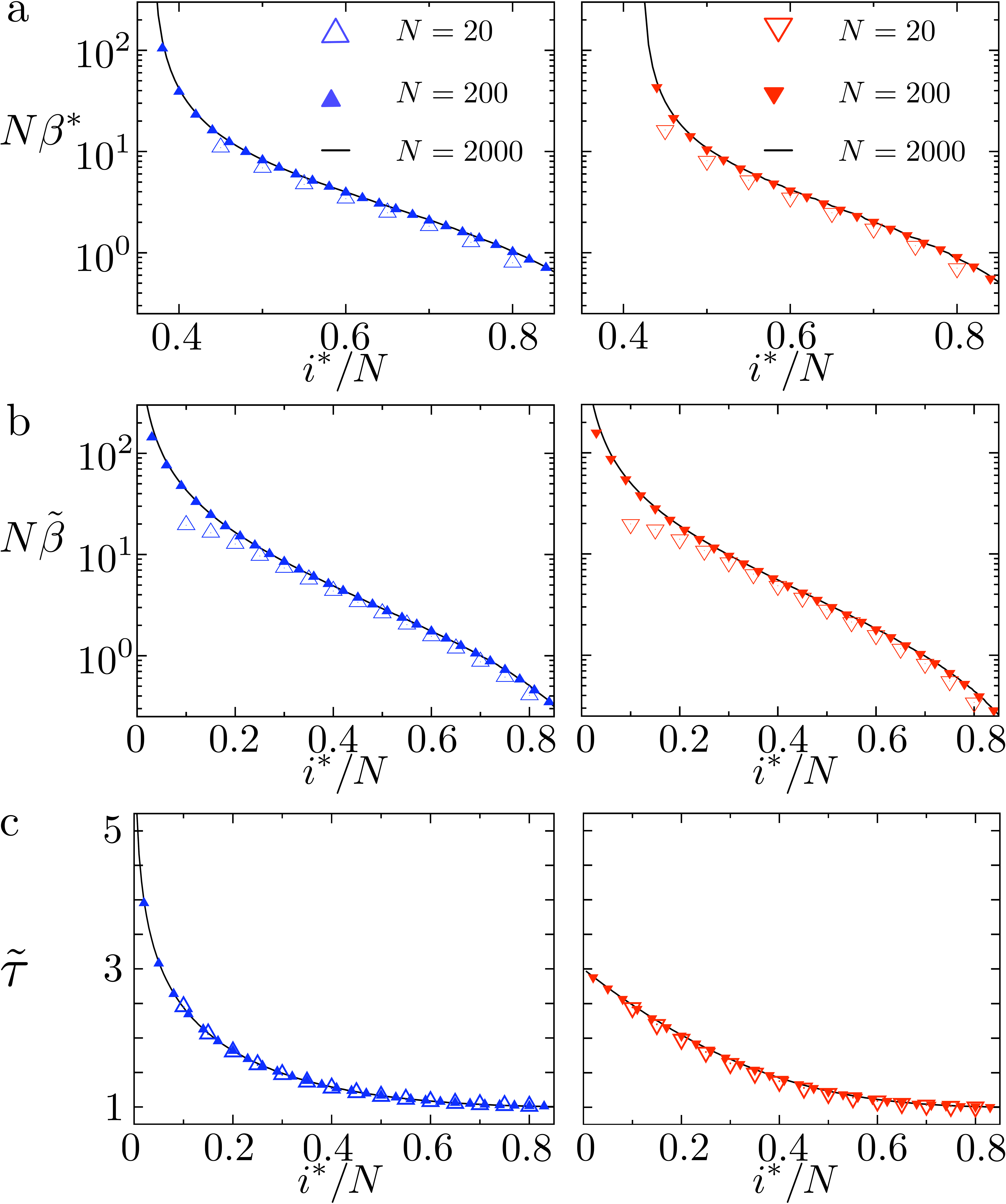}
\caption{
(Color online)
Scaling with system size for the two models with step like asymmetry: 
Parabolic--step model Eq.~\eq{parabolstep} (\fig{hump} (b)) on the left,
constant--step model Eq.~\eq{TranP03} with $c=1/2$ (\fig{hump} (c)) on the right. 
({\bf a}) The threshold value $N\beta^\ast$, defined by $\tau_1^N(\beta^\ast) = \tau_1^N(0)$. 
Note that $\beta\leq1$ permits a minimal value of $i^\ast/N$ only relatively far from zero.  
({\bf b}) $N\tilde\beta$, defined as the bias parameter where the mean exit time $\tau_1^N$ is maximal. 
When plotted against the asymmetry parameter $i^\ast$, 
both models approach a limit curve with growing size $N$.  
This suggests that non--trivial values of $\beta^{\ast}$ and $\tilde\beta$  
can be found for any system size $N$ after appropriate rescaling: 
The asymptotic scaling relations are $\tilde\beta\sim N^{-1}$, and $\beta^\ast\sim N^{-1}$. 
({\bf c}) The maximal increase of the mean exit time (normalized), $\tilde\tau=\tau_1^N(\tilde\beta)/\tau_1^N(0)$, 
quickly approaches a limiting curve with growing $N$. 
This suggests the asymptotic scaling relation $\tilde\tau\sim N^0$.  
Open symbols $N=20$, filled symbols $N=200$, lines $N=2000$.
}
\label{fig:scaling}
\end{figure}

Is there a simpler process with similar characteristics? 
Indeed, we can introduce asymmetry also as a step in the fitness of the two types
in our Moran process. This leads to parabolic transition probabilities with an additional step--like discontinuity, 
\begin{equation}\label{eq:parabolstep}
T_i^\pm=\frac{i(N-i)}{N^2}\left(1\pm\beta\,\Theta[i^\ast-i]\right),
\end{equation}
where $\Theta[x]$ is the step function
($\Theta[x<0]=0$ and $\Theta[x \geq 0]=1$).  
The integer $i^\ast$ is the location of the step.
This process has the fixation probabilities
\begin{align}\label{eq:FPP03_1}
\begin{split}
\phi_i^N=
\begin{cases}
\frac{1}{\phi_1^i}
\frac{
\phi_1^{i^{\ast}}
}
{
\phi_1^{i^{\ast}}(N-i^\ast)\gamma^{i^\ast}+1
}\,\,&\text{if}\,\,i\leq i^\ast,
\\
\vspace{-0.2cm}
\\
\frac{
\phi_1^{i^{\ast}}(i-i^\ast)\gamma^{i^{\ast}}+1
}
{
\phi_1^{i^{\ast}}(N-i^\ast)\gamma^{i^\ast}+1
}\,\,&\text{if}\,\,i\geq i^\ast,
\end{cases}
\end{split}
\end{align}
where $\phi_1^{k}=(1-\gamma)/(1-\gamma^{k})$ is the probability to get from $1$ to $k$, and $\gamma=(1-\beta)/(1+\beta)$. 
Note that this general formula reduces to the standard fixation probability for constant 
fitness in the case of $i^{\ast}=N$, cf.~Eq.~\ref{stdfix}. 
For weak bias, $\beta\ll1/N$, we have $\gamma\approx1-2\beta$, as well as 
\begin{align}\label{eq:FPP03_2}
\begin{split}
\phi_i^N\approx\,&\frac{i}{N}\\&+\frac{\beta}{N^2}
\begin{cases}
i \left[(N (1+2 i^{\ast}-i) - i^{\ast}(1 + i^{\ast})\right] \,\,&\text{if}\,\,i\leq i^{\ast}, 
\\
(N-i) i^{\ast} (1 + i^{\ast})   \,\,&\text{if}\,\,i> i^{\ast}.
\end{cases}
\end{split}
\end{align}
$\phi_i^N$ increases with $\beta$ in this approximation, whereas $\gamma$ decreases with $\beta$. 
Hence, the mean exit time can also increase in an appropriate parameter range. 
The average delay of the absorption is rather high in this case, cf.~\fig{hump} (b), where it is $10 \%$. 
\fig{scaling} (c) illustrates that even a delay of 400\% is possible, 
but this delay decreases with increasing $i^{\ast}$. 

An even simpler model with stochastic slowdown is the constant--step process 
\begin{align}\label{eq:TranP03}
\begin{split}
T_i^\pm=
 c\,(1\pm\beta\,\Theta[i^\ast-i])\,\,&\text{if}\,\,\,\,0<i<N, 
\end{split}
\end{align}
and $T_0^\pm=T_N^\pm = 0$, with $i^{\ast}\leq N$, and the constant $c$ chosen such that $T_i^+ + T_i^-\leq1$. 
Clearly, the fixation probability of this process obeys Eqs.~\eq{FPP03_1} and \eq{FPP03_2}.  
Then, the remaining sums can be expressed by means of the exact form of $\phi_i^N$,
respecting that $1/T_l^+$ only gives contributions different from $1/c$ if $l\leq i^{\ast}$.
The conditional mean exit time $\tau_1^N$ can now be written in the form 
\begin{align}\label{eq:METP03_2}
\begin{split}	
\tau_1^N =&
	\,\frac{\phi_1^N}{c}
	\sum\limits_{k=1}^{i^{\ast}}
	\sum\limits_{l=1}^{k}
	\frac{\gamma^{k-l}(1+\gamma)}
	{2\phi_1^l}
	\\
	&+\frac{\phi_1^N}{c}
	\sum\limits_{k=i^{\ast}+1}^{N-1}
	\sum\limits_{l=1}^{i^{\ast}}
	\frac{\gamma^{i^{\ast}-l}(1+\gamma)}
	{2\phi_1^l}
	\\
	&+\frac{\phi_1^N}{c}\sum\limits_{k=i^{\ast}+1}^{N-1}\sum\limits_{l=i^{\ast}}^{k-1}\!\left[ (k-l)\gamma^{i^{\ast}}\!+\!\frac{1}{\phi_1^{i^{\ast}}}\right]\!.
\end{split}
\end{align}
With $\gamma\approx1-2\beta$ and Eq.~\eq{FPP03_2} this leads to
\begin{align}\label{eq:METP03_4}
	\tau_1^N\approx\frac{N^2-1}{6\,c}+\frac{(N - i^{\ast})(N -1-i^{\ast})  i^{\ast} (1 + i^{\ast})}{3N\,c}\,\beta.
\end{align}
The constant contribution is that of the homogenous random walk. 
The correction linear in $\beta$ is always greater than or equal to zero, i.e.~within the range of this approximation it just adds a positive value to the symmetric part. 
Also note that $\tau_1^N(\beta=0,i^{\ast},c)$ serves as an upper bound for the mean exit time if $i^{\ast}\geq N-1$. 
Hence, below a certain threshold of the bias, $\tau_1^N$ is always greater than or equal to the homogenous random walk between absorbing boundaries. 
This is surprising as the process defined by Eq.~\eq{TranP03} fulfills $T_i^+\geq T_i^-$, and thus never gives a disadvantage to movement towards the boundary $i=N$. 
Moving into the direction of $N$ is always at least as likely as moving into the opposite direction in this setup. 
In this particular process, the stochastic slowdown can be quite large, cf.\ \fig{hump} (c) and \fig{scaling} (c). 

What is the effect of system size on this stochastic slowdown? 
Let $\beta^{\ast}$ denote the upper bound of the parameter $\beta$ for which 
$\tau_1^N(\beta) > \tau_1^N(0)$, which is the parameter range in which slowdown can be observed. 
Additionally, with $\tilde \beta$ we denote the parameter value of maximal slowdown of the exit time $\tau_1^N$. 
They change with $N$ and $i^{\ast}$ in both models with a step--like asymmetry, Eqs.~\eq{parabolstep} and \eq{TranP03}. 
The expansions linear in $\beta$ are valid if $N\beta\ll1$ \cite{taylor:2004wv,antal:2006aa,altrock:2009nj}. 
In \fig{scaling} (a) and (b) we show that with increasing system size $N$, 
the quantities $N\tilde\beta(i^{\ast})$ and $N \beta^{\ast}(i^{\ast})$ approach limiting curves if $\beta$ is rescaled appropriately. 
Thus, stochastic slowdown does not rely on small system size, but $\beta^\ast$ and $\tilde\beta$ asymptotically scale as $N^{-1}$. 
However, the maximal relative increase of the mean exit time itself, $\tilde\tau=\tau_1^N(\tilde\beta)/\tau_1^N(0)$, does not scale with system size, 
$\tilde\tau\sim N^0$, as illustrated in \fig{scaling} (c).

\subsection{Directed mutations}\label{ssec:directed}

To stress the generality of the effect of stochastic slowdown in asymmetric birth--death processes we briefly discuss a model with directed mutations. 
Fitness does not need to be position/state dependent to observe stochastic slowdown in population genetics. 
As above we consider two types, $A$ and $B$, in a population of size $N$, both having the same reproductive fitness. 
In one reproduction step of this Moran process, type $B$ mutates to type $A$ with a probability $\mu$, back--mutations are excluded. 
This introduces asymmetry in the transition rates, 
\begin{align}
	T_i^+&=   \left(\frac{i}{N} +\mu\, \frac{N-i}{N}\right) \frac{N-i}{N}\label{eq:Tmutplus},\\
	T_i^-&= \left(\frac{N-i}{N}(1-\mu)\right) \frac{i}{N} \label{eq:Tmutminus},
\end{align}
where $i$ is the abundance of $A$. 
Obviously, $T_N^\pm=T_0^-=0$, but with directed mutations we have $T_0^+\geq0$. 
The process has one absorbing boundary. 
The ratio of the transition probabilities is $T_m^-/T_m^+\approx1-\mu N/m$, for mutation rates $\mu\ll 1/N^2$. 
For larger $\mu$, the dependence on the inverse mutation rate makes the calculation of an approximation of Eq.~\eq{CtimeOne} unwieldy. 
As $\mu$ increases we expect that $A$ has an advantage during reproduction and hence, the conditional fixation time 
(that a single mutant takes over before going temporarily extinct)
should decrease.
Nevertheless, we observe an increase in the value of $\tau_1^N$, see \fig{hump} (d).  
The time shows a maximum when $\mu$ is close to $N^{-1}$. 

A more general process is given in Appendix \ref{sec:App}. 
There, we derive an expression for the fixation probability in a Wright--Fisher model with directed mutations. 
Although this quantity increases with $\mu$, the associated conditional mean exit time also increases in a certain parameter range, compare \fig{WF}.

\section{State dependent Wright--Fisher process}\label{sec:WF}

\begin{figure}[t]
\includegraphics[angle=0,width=0.475\textwidth]{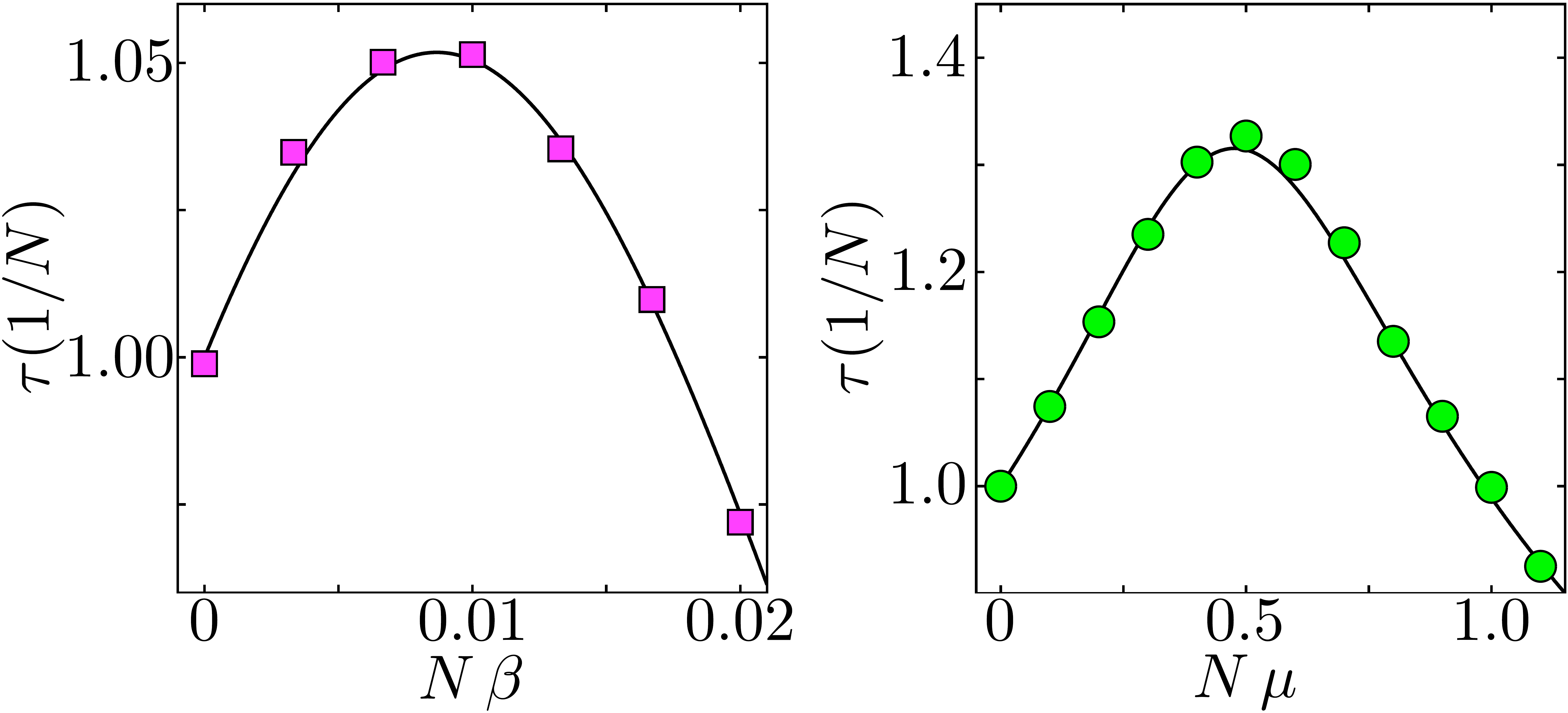}
\caption{
(Color online)
The conditional mean exit time (normalized) for the Wright--Fisher model with $N=1000$, 
as a function of the rescaled bias (selection intensity, mutation rate). 
The line shows the analytical diffusion approximation result Eq.~\eq{WF04}, namely $\tau(N^{-1})/(2N-1)$.  
Symbols are simulation results. 
Left: The state dependent fitness model, Eq.~\eq{WF01} ($2\times10^6$ realizations, $a=-0.1$, $b=N\,|a|$). 
For relatively small bias $\beta$ slowdown is observed. 
Right: The directed mutations model, Eq.~\eq{App01} ($5\times10^5$ realizations). 
Here, a strong slowdown effec can be observed over a wide range of the bias, $N\mu\leq1$. 
This is due to the different nature of the directed mutation process, which has only one absorbing boundary.  
}
\label{fig:WF}
\end{figure}

The phenomenon of stochastic slowdown is not restricted to birth--death processes.  
It also occurs in the Wright--Fisher process that is commonly used in population genetics \cite{ewens:2004qe,imhof:2006aa}. 
Again, we consider a population of two types $A$ and $B$. 
If $i$ is the abundance of $A$, the fitness of each type is $f_A=1+\beta(a i+b)$, and $f_B=1-\beta(a i+b)$, respectively.  
Birth--death processes, such as the Moran model considered above, deal with one reproductive event at a time. 
Now, one time step of the Wright--Fisher process corresponds to one generation where all individuals reproduce: 
In each generation, the $N$ individuals reproduce a large number of offspring proportional to fitness.
The new generation of size $N$ is a random sample from this offspring pool, 
which corresponds to binomial sampling proportional to fitness.
The transition probability to go from $i$ to $j$ $A$ individuals reads \cite{imhof:2006aa}
\begin{align}\label{eq:WF01}
\begin{split}
	T_{i\to j}=\binom{N}{j}\,&
	\!\!
	\left( \frac{i\,f_A}{i\,f_A+(N-i)f_B} \right)^{j}\\
	\times&\left( \frac{(N-i)\,f_B}{i\,f_A+(N-i)f_B} \right)^{N-j}.
\end{split}
\end{align}
For this process, a closed treatment is not possible. 
Apart from simulations, for large $N$ a diffusion approximation leads to analytical results \cite{maruyama:1977ln,buerger:2000aa,ewens:2004qe,traulsen:2006ab,chalub:2006cc}. 
With $x=i/N$, the process is approximately described by the Langevin equation $dx=D_1(x)dt+\sqrt{D_2(x)}dW(t)$, 
where $W(t)$ is the Wiener process with zero mean and autocorrelation $\langle W(t)W(s)\rangle=\min(t,s)$, \cite{kampen:1997xg}. 
The drift term $D_1(x)$ can be written as
\begin{align}\label{eq:Drift01}
	D_1(x)=x(1-x)N\frac{f_A(x)-f_B(x)}{x f_A(x)+(1-x)f_B(x)}. 
\end{align}
For the diffusion term $D_2(x)$ we find 
\begin{align}\label{eq:Diff01}
\begin{split}
	D&_2(x)=x(1-x)\frac{f_A(x)f_B(x)}{\left(x f_A(x)+(1-x)f_B(x)\right)^2}+\frac{D_1^2(x)}{N}.
\end{split}
\end{align}
If the initial fraction of $A$ types is $x_0$, the probability of absorption in $x=1$ (fixation probability) reads
\begin{align}\label{eq:WF02}
	\phi(x_0)=\frac{S(x_0)}{S(1)},
\end{align} 
where 
\begin{align}\label{eq:WF022}
	S(x)=\int_0^{x}dy\exp\left[-\,\int_0^ydz\frac{2D_1(z)}{D_2(z)}  \right].
\end{align} 
If there is no bias, $\beta=0$, we have $f_A(x)=f_B(x)$ and hence $D_1(x)=0$. 
Thus, consistently with the previous section, we obtain $\phi(i/N)=i/N$. 
For sufficiently weak bias, $N\beta\ll1$, we have 
\begin{align}
\label{eq:d1d2}
	\frac{2D_1(z)}{D_2(z)}\approx4N(a\,N\,z+b)\,\beta
\end{align} 
which leads to
\begin{align}\label{eq:WF03}
	\phi(x_0)\approx x_0+\frac{2\,x_0(1-x_0)N[aN(1+x_0)+3b]}{3}\,\beta. 
\end{align}
The conditional mean time this process takes to exit at $x=1$, $\tau(x_0)$, can be obtained from the associated backward Fokker--Planck equation \cite{ewens:2004qe}, 
\begin{align}\label{eq:WF04}
\begin{split}
	\tau(x_0)=N\,\int\limits_{0}^{x_0}dx\,t_1(x,x_0)+N\,\int\limits_{x_0}^{1}dx\,t_2(x,x_0),
\end{split}
\end{align}
where
\begin{align}\label{eq:WF05}
\begin{split}
		t_1(x,x_0)&=\, 2\frac{\phi(x)}{D_2(x)}\frac{1-\phi(x_0)}{\phi(x_0)}S(x)\exp\left[\int\limits_0^x dz\frac{2D_1(z)}{D_2(z)}\right],\\
		t_2(x,x_0)&=2\frac{\phi(x)}{D_2(x)}\left( S(1)-S(x) \right)\exp\left[\int\limits_0^x dz\frac{2D_1(z)}{D_2(z)}\right].
\end{split}
\end{align}
For weak bias Eq.~\eq{d1d2} holds, as well as $S(x)\approx x-2/3Nx^2(aNx+3b)\beta$. 
This results in
\begin{align}\label{eq:WF06}
\begin{split}
	\tau(1/N) \approx&2N(N-1) \ln\left[ \frac{N}{N-1}\right]	\\
	& -\frac{2}{9}(N-1)\left( C_1+C_2\ln \left[ \frac{N-1}{N} \right] \right)\,\beta,
\end{split}
\end{align}
with 
\begin{align}
	C_1\,=\,&a(7N^2+13N+6)+18b, \nonumber\\
	C_2\,=\,&6N(aN(N+2)+3b).  \nonumber
\end{align}
For large $N$, the right hand side of Eq.~\eq{WF06} simplifies, leading to
\begin{align}\label{eq:WF07}
	\tau(1/N)\approx2N-1-a\,\frac{2N^2(N-3)}{9}\,\beta.
\end{align}
Hence, we can predict an increase of $\tau(1/N)$, in the case of state dependent bias with $a<0$, also for the Wright--Fisher process, 
in particular when $A$ always has a fitness advantage over $B$, see \fig{WF}.  
This goes along with the findings for the Moran model in the previous section. 
Thus, the slowdown effect can also be observed in the traditional framework of population genetics, 
where times of fixation (or rather extinction) have been considered typically for constant selection \cite{nei:1973ge,ewens:2004qe}.

\section{Discussion}\label{disc}

This manuscript addresses several stochastic evolutionary processes asking how long an advantageous mutation needs to take over. 
We have first concentrated on birth--death processes which model population dynamics with successive reproductive events, like the Moran process.  
However, the phenomenon of stochastic slowdown is also present in more general Markov processes, 
e.g.~the Wright--Fisher process from population genetics. 
Stochastic slowdown is relevant in the invasion and fixation of beneficial traits with small state dependent selective advantage,
which is typically assumed in evolutionary biology \cite{ohta:1973aa}. 
However, consequences of weak, but non-vanishing selection are hard to reveal in empirical studies, 
as the dynamics are still dominated by random genetic drift and averages over large ensembles 
are necessary. 
Biological examples of weak selection include amino acid substitutions which are only slightly advantageous or deleterious \cite{ohta:1997aa,bustamante:2002aa,fay:2002}.
Weak state dependent fitness changes (such as the thresholds we discuss in our model with step-like asymmetry)
may help explain situations in which a substitution is likely, but takes a very long time. 

Our finding also has applications in evolutionary game theory \cite{nowak:2004aa,nowak:2006pw,cremer:2009njp}: 
When a group of cooperative individuals is eventually driven to exinction by defectors, 
this process may take longer than the corresponding neutral process, although the defectors always have a fitness advantage. 
This observation is closely related to the fact that the conditional fixation time of an advantageous mutation is the same as the conditional fixation time of a deleterious mutation \cite{taylor:2006jt,nei:1973ge}. 

To sum up, we have shown that an asymmetric bias in a random walk, which is generic in population genetics, 
can lead to a counterintuitive observation that an advantageous mutant needs longer to take over the population than a neutral mutant in the same system. 
This is a property of weakly biased systems, i.e.~weak selection, and is recovered for any system size if the intensity of selection is rescaled with $N^{-1}$. 
The relative maximal increase in time itself is independent of the system size. 

Especially in the state dependent Moran or Wright--Fisher process, 
this can have a crucial impact on macroscopic observable quantities.

\section*{Acknowledgement}
Financial support by the Emmy-Noether program of the DFG is gratefully acknowledged. 
We also thank an anonymous referee for helpful suggestions. 

\newpage

\appendix

\section{State dependent Wright--Fisher process with directed mutations}\label{sec:App}

Consider a finite population of size $N$, which consists of two types $A$ and $B$. 
Both types have the same reproductive rate, which is set to one. 
In one generation, each type produces a large number of identical offspring proportional to its abundance.
Additionally, a directed mutation from $B$ to $A$ can occur with probability $\mu$. 
The next generation of size $N$ is a random sample from the offspring pool. 
The transition matrix reads
\begin{align}\label{eq:App01}
\begin{split}
	T_{i\to j}=\binom{N}{j}\,&
	\!\!
	\left(  \frac{i}{N}+\mu\frac{N-i}{N}\right)^{j}
	\left(\frac{N-i}{N}(1-\mu) \right)^{N-j}.
\end{split}
\end{align}
The conditional moments of this Markov chain are given by \cite{ewens:2004qe}
\begin{align}\label{eq:App02}
	\mathcal M_n(i)=\,\sum\limits_{j=0}^{N}(j-i)^n\,T_{i\to j}.
\end{align}
In a diffusion approximation we rescale the state space as $x=i/N$, and the timescale as $\Delta t=1/N$,  
such that for large $N$ the process is well described by the first two moments,
\begin{align}
	D_k(x)=\frac{N}{N^k}\mathcal M_k(i),\label{eq:App03}
\end{align}
$k=1,2$. 
For the given Markov chain Eq.~\eq{App01}, 
the drift and diffusion terms read
\begin{align}
	D_1(x)&=\mu\,N\,(1-x),\label{eq:App04}\\
	D_2(x)&=(1-x)\left( (1-x)(N-1)\mu^2+(1-2x)\mu+x \right)\label{eq:App05}. 
\end{align}
Next, we derive a closed expression for the probability that the process exits at $x=1$ without hitting the non--absorbing boundary $x=0$ first, 
starting from $x_0$, $\phi(x_0)$, Eq.~\eq{WF02}. 
The general expressions Eqs.~\eq{WF02}, \eq{WF022}, as well as Eqs.~\eq{WF04}, \eq{WF05} hold. 
However, due to the different nature of this process, where only one absorbing boundary at $x=1$ exists, these quantities have a slightly different meaning. 

With $2\,D_1(x)/D_2(x)=2N\mu/\widetilde D_2(x)$, where 
\begin{align}\label{eq:App055}
	\widetilde D_2(x)=(1-x)(N-1)\mu^2+(1-2x) \mu +x, 
\end{align}	
we obtain
\begin{align}\label{eq:App06}
\begin{split}
	I_1(y)=\int\limits^ydz\frac{2\,D_1(z)}{D_2(z)}=-\nu\,\ln \widetilde D_2(y)
\end{split}
\end{align}
with 
\begin{align}\label{eq:App07}
\begin{split}
	\nu=\frac{2N\mu}{\mu((N-1)\mu+2)-1}.
\end{split}
\end{align}
Now, with \[I_2(y)=\exp \{-(I_1(y)-I_1(0))\}=\bigg(\frac{\widetilde D_2(y)}{D_2(0)}\bigg)^{\nu},\] we can calculate the second integral in Eq.~\eq{WF022},
\begin{align}\label{eq:App08}
\begin{split}
	S(x)=&\,\int\limits_{0}^{x}dy\,I_2(y)\\
	=&\,\frac{1}{D_2^\nu(0)}\frac{\widetilde D_2^{\nu+1}(x)-\widetilde D_2^{\nu+1}(0)}{\mu (2 - \mu + N (2 + \mu))-1}.
\end{split}
\end{align}
Hence, the fixation probability, Eq.~\eq{WF02}, reads
\begin{align}\label{eq:App09}
\begin{split}
	\phi(x_0)=\frac{\widetilde D_2^{\nu+1}(x_0)-\widetilde D_2^{\nu+1}(0)}{\widetilde D_2^{\nu+1}(1)-\widetilde D_2^{\nu+1}(0)}.
\end{split}
\end{align}
As $\widetilde D_2(0)=((N-1)\mu+1)\mu$, $\widetilde D_2(1)=1-\mu$, and $\lim_{\mu\to0}\widetilde D_2(x_0)=x_0$, we have $\lim_{\mu\to0}\phi(x_0)=x_0$. 
Up to first order in mutation rate, we see that $\phi(x_0)$ increases with increasing bias, 
\begin{align}\label{eq:App10}
\begin{split}
	\phi(x_0)\approx x_0 - \left(2N\,x_0 \ln x_0\right)\,\mu.
\end{split}
\end{align}
With these expressions the conditional mean exit time, Eq.~\eq{WF04}, can be tackled as well. 
However, we do not address the conditional mean exit time analytically, as its explicit form is elaborate and does not lead to further insight. 
From a numerical solution (Eq.~\eq{WF04}) and from simulations the mean exit time of a single mutant, $\tau(1/N)$, as a function of $\mu$ is shown in \fig{WF}.

\bibliographystyle{h-physrev3}

\end{document}